\documentclass[runningheads]{llncs}

\usepackage{graphicx}
\usepackage{amsmath}
\usepackage{amssymb}
\usepackage{comment}

\begin{document}

\title{Logic-Driven Cybersecurity: A Novel Framework for System Log Anomaly Detection using Answer Set Programming}

\author{Fang Li\inst{1} \and Fei Zuo\inst{2} \and Gopal Gupta\inst{3}}

\institute{Oklahoma Christian University \and
University of Central Oklahoma \and
University of Texas at Dallas}

\maketitle

\begin{abstract}
This study explores the application of Answer Set Programming (ASP) for detecting anomalies in system logs, addressing the challenges posed by evolving cyber threats. We propose a novel framework that leverages ASP's declarative nature and logical reasoning capabilities to encode complex security rules as logical predicates. Our ASP-based system was applied to a real-world Linux system log dataset, demonstrating its effectiveness in identifying various anomalies such as potential brute-force attacks, privilege escalations, frequent network connections from specific IPs, and various system-level issues. Key findings highlight ASP's strengths in handling structured log data, rule flexibility, and event correlation. The approach shows promise in providing explainable alerts from real-world data. This research contributes to computer forensics by demonstrating a logic-based paradigm for log analysis on a practical dataset, opening avenues for more nuanced and adaptive cyber intelligence systems. 

\end{abstract}


\section{Introduction}

In the ever-evolving landscape of cybersecurity, the detection of anomalies in system logs remains a critical challenge. As cyber threats grow in sophistication, traditional methods of log analysis often struggle to identify subtle patterns of malicious activity, leaving organizations vulnerable to APTs (Advanced Persistent Threats), insider attacks, and novel exploitation techniques. The sheer volume and complexity of system logs generated by modern IT infrastructures further compound this challenge, making it increasingly difficult for security analysts to effectively identify and respond to potential threats in a timely manner.

Existing approaches to log analysis and anomaly detection, such as statistical methods, machine learning algorithms, and rule-based systems, each have their strengths but also significant limitations. Statistical methods may fail to capture complex, multi-stage attack patterns. Machine learning approaches, while powerful, often lack explainability and require large, labeled datasets for training. Traditional rule-based systems, though interpretable, can be inflexible and struggle with incomplete or ambiguous information.

This paper proposes a novel approach to anomaly detection in system logs using Answer Set Programming (ASP), a declarative programming paradigm based on stable model semantics. While simple threshold-based approaches could detect individual anomaly types, ASP provides unique technical advantages for cybersecurity log analysis: (1) Non-monotonic reasoning enables detection of complex multi-stage attacks where absence of expected events is significant, (2) Default reasoning handles incomplete log data gracefully without requiring complete information, (3) Declarative rules allow security experts to encode complex policies without algorithmic implementation details, and (4) The logical structure provides explainable alerts with clear reasoning chains crucial for security analysis and forensic investigation.

Our ASP-based framework encodes a wide range of security rules and heuristics as logical predicates. We demonstrate the effectiveness of this approach by applying it to a publicly available real-world Linux system log dataset, showcasing its ability to detect various anomalies. The remainder of this paper is organized as follows: Section 2 provides background on ASP and its applications in cybersecurity. Section 3 details our methodology, including the data preprocessing, the design of the ASP-based anomaly detection system tailored to the dataset, and the experimental setup. Section 4 presents the results of our experiments on this dataset and discusses their implications. Section 5 addresses current limitations and challenges of the approach. Finally, Section 6 concludes the paper and outlines directions for future research, including potential integrations with machine learning techniques and the development of real-time processing capabilities.

By introducing this logic-based paradigm for log analysis, we aim to contribute to the field of computer forensics and cybersecurity, offering new possibilities for more nuanced, adaptive, and explainable anomaly detection systems. Our work paves the way for future research into hybrid approaches that combine the strengths of logical reasoning with other computational techniques, potentially revolutionizing how organizations detect and respond to cyber threats.

\section{Background and Related Work}

\subsection{Answer Set Programming} \label{sec:asp}   

Logic programming \cite{abiteboul1995foundations} has been applied to many areas such as fault diagnosis, databases, planning, natural language processing, knowledge representation and reasoning. During decades of exploration, researchers have developed various semantics for solving different reasoning tasks. Among those semantics, the stable model semantics based answer set programming (ASP) \cite{MT5} paradigm is popular for knowledge representation and reasoning as well as for solving combinatorial problems. Though computing ASP programs is considered to be NP-hard, there are a lot of ASP solvers (e.g., CLINGO \cite{gebser2014clingo}, DLV \cite{dlv}, s(CASP) \cite{arias2018constraint}) that can compute stable models of an ASP program efficiently.

Answer Set Programming (ASP) is a declarative paradigm that extends logic programming with negation-as-failure. ASP is a highly expressive paradigm that can elegantly express complex reasoning methods, including those used by humans, such as default reasoning, deductive and abductive reasoning, counterfactual reasoning, constraint satisfaction~\cite{baral,gelfond2014knowledge}.
ASP supports better semantics for negation ({\it negation as failure}) than does standard logic programming and Prolog. An ASP program consists of rules that look like Prolog rules. The semantics of an ASP program {$\Pi$} is given in terms of the answer sets of the program \texttt{ground($\Pi$)}, where \texttt{ground($\Pi$)} is the program obtained from the substitution of elements of the \textit{Herbrand universe} for variables in $\Pi$~\cite{baral}.
Rules in an ASP program are of the form shown as below (Rule \ref{rule1}):
\begin{equation}
\label{rule1}
    p \; :- \; q_1, \; ..., \; q_m, \; \textbf{not} \; r_1, \; ..., \; \textbf{not} \; r_n.
\end{equation} 
\noindent where $m \geq 0$ and $n \geq 0$. Each of \texttt{p} and \texttt{q$_i$} ($\forall i \leq m$) is a literal, and each \texttt{not r$_j$} ($\forall j \leq n$) is a \textit{naf-literal} (\texttt{not} is a logical connective called \textit{negation-as-failure} or \textit{default negation}). The literal \texttt{not r$_j$} is true if proof of {\tt r$_j$} \textit{fails}. Negation as failure allows us to take actions that are predicated on failure of a proof. 
Thus, the rule {\tt r :- not s.} states that {\tt r} can be inferred if we fail to prove {\tt s}. 
Note that in Rule \ref{rule1}, {\tt p} is optional. Such a headless rule is called a constraint, which states that conjunction of {\tt q$_i$}'s and \texttt{not r$_j$}'s should yield \textit{false}. Thus, the constraint {\tt :- u, v.} states that {\tt u} and {\tt v} cannot be both true simultaneously in any model of the program (called an answer set).

The declarative semantics of an Answer Set Program \texttt{P} is given via the Gelfond-Lifschitz transform~\cite{baral,gelfond2014knowledge} in terms of the answer sets of the program \texttt{ground($\Pi$)}. 
More details on ASP can be found elsewhere~\cite{baral,gelfond2014knowledge}.

In general, the key features of ASP include:

\begin{enumerate}
    \item Declarative Semantics: Programs describe the problem rather than the solution algorithm.
    \item Non-monotonic Reasoning: ASP supports default reasoning and the ability to draw conclusions based on the absence of information.
    \item Efficient Solvers: Modern ASP solvers can handle large-scale problems efficiently.
\end{enumerate}

\textbf{Example of ASP Rule and Reasoning:}

Consider the following ASP rule:

\begin{verbatim} 
    can_fly(X) :- bird(X), not abnormal(X). 
    abnormal(X) :- penguin(X). 
    bird(tweety). 
    penguin(polly). 
\end{verbatim}

This program states that:
\begin{itemize}
    \item \texttt{can\_fly(X)}: An entity \texttt{X} can fly if it is a \texttt{bird} and it is not \texttt{abnormal}.
    \item \texttt{abnormal(X)}: An entity \texttt{X} is abnormal if it is a \texttt{penguin}.
    \item \texttt{bird(tweety)}: \texttt{tweety} is a bird.
    \item \texttt{penguin(polly)}: \texttt{polly} is a penguin.
\end{itemize}

The answer set for this program would be:

\begin{verbatim}
    bird(tweety). penguin(polly). 
    abnormal(polly). can_fly(tweety).
\end{verbatim}

This demonstrates how ASP uses \texttt{negation-as-failure}. Since there is no evidence that \texttt{tweety} is \texttt{abnormal}, the rule \texttt{can\_fly(tweety)} is inferred. However, since \texttt{polly} is a {penguin}, \texttt{abnormal(polly)} is inferred, and therefore \texttt{can\_fly(polly)} is not inferred.

\subsection{System Logs}

System logs are a rich source of information generated by various IT components, including operating systems, network devices, and applications. They contain a wealth of data about system events, user activities, and configuration changes. System logs play a critical role in monitoring, troubleshooting, and ensuring the security and stability of IT infrastructures. Analyzing system logs can help security analysts identify potential security threats, policy violations, and system performance issues. However, the sheer volume and heterogeneity of system logs pose challenges in identifying anomalies and discerning significant patterns from the noise. As a result, sophisticated anomaly detection methods, such as those based on Answer Set Programming (ASP), are essential for efficiently and accurately analyzing system logs to enhance cybersecurity defenses and improve overall system performance.

\subsection{ASP in Anomaly Detection}

ASP has been increasingly recognized as a valuable tool for anomaly detection~\cite{bellusci2022modelling} and cybersecurity applications~\cite{rezvani2019analyzing}. The declarative, rule-based nature of ASP enables security analysts to concisely encode complex patterns and relationships between entities~\cite{li2025exploring}. This makes it possible to detect subtle anomalies that might be difficult to identify using traditional rule-based or statistical techniques. Furthermore, the interpretability of ASP allows for clear, easily understood explanations of detected anomalies, facilitating a better understanding of the underlying causes and contributing to a more efficient response.

ASP offers a unique combination of features that make it well-suited for anomaly detection in system logs:
\begin{itemize}
    \item \textbf{Declarative Knowledge Representation:} Security rules and patterns can be expressed in a high-level, logical format, making it easier for security experts to encode complex policies.
    \item \textbf{Non-monotonic Reasoning:} ASP can handle incomplete information and make default assumptions, crucial when dealing with partial or corrupted log data.
    \item \textbf{Expressiveness:} Complex relationships between different types of log entries can be easily represented, allowing for sophisticated pattern matching.
    \item \textbf{Explainability:} The logical structure of ASP programs provides clear reasoning chains, helping analysts understand why certain events were flagged as anomalous.
\end{itemize}

\section{Methodology}

This study aims to evaluate the effectiveness of Answer Set Programming (ASP) in detecting anomalies in system logs and compare it with traditional anomaly detection techniques. The methodology consists of the following steps:

\subsection{Anomaly Pattern Definition} Based on the analysis of the dataset, we defined and refined several anomaly patterns using ASP rules. These rules target specific event types and behaviors observable in the dataset.

\subsubsection{Brute Force Attempts} Detects multiple failed login attempts from the same IP address within a short time window. This is a common attack technique used to gain unauthorized access to a system.

\textbf{ASP Rule:} 
\begin{verbatim} 
    #const login_time_window = 600. % 10 minutes
    #const failed_login_threshold = 5. % 5 failed attempts
    
    potential_brute_force(IP, Time) :- 
        login_attempt(IP, Time, failed, _),
        failed_count(IP, Time, Count),
        Count >= failed_login_threshold.
    
    failed_count(IP, Time, Count) :- 
        login_attempt(IP, Time, failed, _), 
        Count = #count{T : login_attempt(IP, T, failed, _), 
        T >= Time - login_time_window, T <= Time}. 
\end{verbatim}

\textbf{Explanation:}
\begin{itemize}
    \item \texttt{login\_time\_window}: Defines the time window (in seconds) within which failed login attempts are counted. A value of 600 seconds (10 minutes) is used in this study.
    \item \texttt{failed\_login\_threshold}: Defines the minimum number of failed login attempts required to trigger a brute-force alert. A value of 5 failed attempts is used in this study.
    \item The \texttt{potential\_brute\_force} rule identifies an IP address (\texttt{IP}) and a specific time (\texttt{Time}) if the number of failed login attempts from that IP within the \texttt{login\_time\_window} meets or exceeds the \texttt{failed\_login\_threshold}.
    \item The \texttt{failed\_count} rule calculates the number of failed login attempts from a given IP address within the specified time window.
\end{itemize}

\textbf{Rationale:}

This rule is based on the observation that brute-force attacks typically involve a large number of failed login attempts from the same IP address within a short period. By setting appropriate values for \texttt{login\_time\_window} and \texttt{failed\_login\_\\threshold}, we can effectively detect potential brute-force attacks while minimizing false positives.

\textbf{Potential Refinements:}
\begin{itemize}
    \item Incorporate successful login attempts after a series of failures as part of the brute-force pattern.
    \item Use a sliding time window to more accurately capture the temporal dynamics of brute-force attacks.
    \item Adjust the thresholds based on the characteristics of the system and the expected user behavior.
\end{itemize}

\subsubsection{Privilege Escalation} Detects instances where a non-root user opens a session with root privileges (UID 0). This can indicate unauthorized attempts to gain elevated access to the system.

\textbf{ASP Rule:} 
    \begin{verbatim} potential_privilege_escalation(User, Timestamp) :- 
        session(User, UID, opened, Timestamp), 
        UID == 0, % Session opened as root 
        User != "root". 
\end{verbatim}

\textbf{Explanation:}

This rule identifies a potential privilege escalation if a user other than "root" opens a session with a user ID (UID) of 0, which typically corresponds to the root user.

\textbf{Rationale:}

This rule is based on the principle that privilege escalation should be a rare event and should only occur through authorized channels. By flagging instances where a non-root user directly assumes root privileges, we can detect potential security breaches or misconfigurations.

\textbf{Potential Refinements:}
\begin{itemize}
    \item Incorporate information about the commands executed during the session to determine if the privilege escalation was legitimate or malicious.
    \item Track the history of user privileges to detect sudden or unexpected changes in access rights.
    \item Use machine learning to learn the normal patterns of privilege escalation and flag deviations from these patterns.
\end{itemize}

\subsubsection{Network Anomalies (Frequent Connections)} 

Identifies IP addresses that establish an unusually high number of network connections, potentially indicating scanning or other malicious network activity.

\textbf{ASP Rule:} 
\begin{verbatim} 
    #const connection_threshold = 10.
    
    network_anomaly(IP) :- frequent_connections(IP).
    
    frequent_connections(IP) :- network_connection(IP, _),
        #count{T : network_connection(IP, T)} >= connection_threshold. 
\end{verbatim}

\textbf{Explanation:}
\begin{itemize}
    \item \texttt{connection\_threshold}: Defines the minimum number of network connections required to trigger a network anomaly alert. A value of 10 connections is used in this study.
    \item The \texttt{network\_anomaly} rule identifies an IP address (\texttt{IP}) as anomalous if it establishes a high number of connections, as determined by the \texttt{frequent\_connections} rule.
    \item The \texttt{frequent\_connections} rule counts the number of network connections from a given IP address and flags it as anomalous if the count meets or exceeds the \texttt{connection\_threshold}.
\end{itemize}

\textbf{Rationale:}

This rule is based on the observation that malicious actors often perform network scanning or other automated activities that involve establishing a large number of connections from a single IP address.

\textbf{Potential Refinements:}
\begin{itemize}
    \item Incorporate information about the destination IPs and ports to identify specific types of network scanning or attacks.
    \item Use a time window to count the number of connections within a specific period.
    \item Adjust the \texttt{connection\_threshold} based on the characteristics of the network and the expected traffic patterns.
\end{itemize}

\subsubsection{Account Anomalies} Flags the creation of new administrator accounts or the reactivation of potentially dormant accounts (requires \texttt{dormant\_account} facts to be effective for reactivation).

\textbf{ASP Rule:} 
\begin{verbatim}
    account_anomaly(Action, User, Time) :- 
        account_action(Action, User, Time), Action == create_admin.
    
    account_anomaly(Action, User, Time) :- 
        account_action(Action, User, Time), 
        Action == reactivate, dormant_account(User). 
\end{verbatim}

\textbf{Explanation:}
\begin{itemize}
    \item This rule identifies account anomalies based on the \texttt{account\_action} and \texttt{dormant\_account} facts.
    \item It flags the creation of new administrator accounts (\texttt{Action == create\_admin}).
    \item It flags the reactivation of dormant accounts (\texttt{Action == reactivate}) if the account is also known to be dormant (\texttt{dormant\_account(User)}).
\end{itemize}

\textbf{Rationale:}

The creation of new administrator accounts and the reactivation of dormant accounts are often associated with malicious activity. By flagging these events, we can identify potential insider threats or unauthorized access attempts.

\textbf{Potential Refinements:}
\begin{itemize}
    \item Incorporate information about the time of day, location, and other characteristics of the account action to identify unusual or suspicious behavior.
    \item Use machine learning to learn the normal patterns of account creation and reactivation and flag deviations from these patterns.
    \item Implement a more robust method for identifying dormant accounts, such as tracking the last login time and flagging accounts that have not been used for a long period.
\end{itemize}

\subsubsection{Klogind Brute Force} Detects multiple klogind authentication failures from the same IP.

\textbf{ASP Rule:} 
\begin{verbatim} 
    #const klogind_failure_threshold = 5.
    
    potential_klogind_brute_force(IP) :- 
        klogind_auth_failure(IP, _), 
        #count{T : klogind_auth_failure(IP, T)} >= 
            klogind_failure_threshold. 
\end{verbatim}

\textbf{Explanation:}
\begin{itemize}
    \item This rule identifies a potential brute-force attack on the \texttt{klogind} service if the number of authentication failures from a given IP address meets or exceeds the \texttt{klogind\_failure\_threshold}.
\end{itemize}

\textbf{Rationale:}

Multiple authentication failures from the same IP address within a short period can indicate a brute-force attack attempting to guess valid credentials for the \texttt{klogind} service.

\textbf{Potential Refinements:}
\begin{itemize}
    \item Incorporate a time window to count the number of failures within a specific period.
    \item Adjust the \texttt{klogind\_failure\_threshold} based on the characteristics of the system and the expected user behavior.
\end{itemize}

\subsubsection{GDM Authentication Failures} 

Detects multiple GDM (GNOME Display Manager) authentication failures.

\textbf{ASP Rule:} 
\begin{verbatim} 
    #const gdm_failure_threshold = 3.
    
    potential_gdm_brute_force(Time) :- gdm_auth_failure(Time), 
        #count{T : gdm_auth_failure(T)} >= gdm_failure_threshold. 
\end{verbatim}

\textbf{Explanation:}
\begin{itemize}
    \item This rule identifies a potential brute-force attack on the GDM service if the number of authentication failures exceeds the \texttt{gdm\_failure\_threshold}.
\end{itemize}

\textbf{Rationale:}

Multiple authentication failures for GDM can indicate a brute-force attack attempting to gain access to the graphical user interface.

\textbf{Potential Refinements:}
\begin{itemize}
    \item Incorporate IP address information, if available, to identify the source of the attack.
    \item Adjust the \texttt{gdm\_failure\_threshold} based on the characteristics of the system and the expected user behavior.
\end{itemize}

\subsubsection{System Issues} 

Flags various system-level issues observed in the logs.

\textbf{ASP Rules:} 
\begin{verbatim}                    
    system_issue(logrotate_abnormal_exit, Time) :- 
        logrotate_abnormal_exit(Time).
    
    system_issue(named_soa_error, Time) :- 
        named_soa_error(Time).
    
    system_issue(xinetd_connection_reset, Time) :- 
        xinetd_connection_reset(Time). 
\end{verbatim}

\textbf{Explanation:}
\begin{itemize}
    \item These rules simply flag the occurrence of specific system-level events as potential issues.

\end{itemize}

\textbf{Rationale:}

These events can indicate underlying problems with the system's configuration or operation.

\textbf{Potential Refinements:}
\begin{itemize}
    \item Incorporate additional information about the context of these events to determine their severity and potential impact.
    \item Correlate these events with other anomalies to identify more complex system-level problems.
\end{itemize}

\subsection{System Log Encoding} 

The primary data source for this study is the Linux logs from a publicly available dataset\cite{zhu2023loghub}. The data were collected from a Linux server over a period of 263.9 days.
The dataset consists of 2,000 structured log entries, each with descriptive fields such as \texttt{Month}, \texttt{Date}, \texttt{Time}, \texttt{Component}, \texttt{PID}, \texttt{Content}, \texttt{EventId}, and \texttt{EventTemplate}.
The distribution of log entries across different components and \texttt{EventIds} is not uniform, reflecting the diverse activities and events occurring on the system.
\begin{itemize}
    \item \textbf{Components:} The \texttt{Component} column indicates the source of the log message (e.g., \texttt{sshd(pam\_unix)}, \texttt{su(pam\_unix)}, \texttt{ftpd}, \texttt{kernel}, \texttt{systemd}). The frequency of different components varies, with \texttt{sshd(pam\_unix)} and \texttt{ftpd} being among the most common.
    \item \textbf{EventIds:} The \texttt{EventId} column provides a coarse-grained categorization of the log event (e.g., \texttt{E16} for authentication failure, \texttt{E29} for connection from, \texttt{E102} for session opened). The distribution of EventIds reflects the prevalence of different types of events in the logs.
    \item \textbf{Content:} The \texttt{Content} column contains the raw log message, which often includes valuable information such as IP addresses, usernames, file paths, and error messages. However, this information is often unstructured and requires parsing to extract meaningful features.
\end{itemize}

\textbf{Data Preprocessing Steps:}

To prepare the data for ASP-based analysis, a Python script
was developed. This script performs the following key steps:

\begin{enumerate} 
\item \textbf{Loading Data:} Reads the 
data file.
\item \textbf{Timestamp Conversion:} Combines the \texttt{Month}, \texttt{Date}, and \texttt{Time} fields into a standardized datetime object using the \texttt{datetime} library. The year is assumed to be the current year for simplicity. The resulting datetime object is then converted to a Unix timestamp (integer seconds) for temporal reasoning in ASP. 
\item \textbf{Information Extraction:} Uses regex 
to extract critical information from the \texttt{Content} field, such as IP addresses and usernames, based on patterns associated with specific \texttt{EventId}'s or log components. For example,
\begin{itemize}
   \item \textbf{IP Address Extraction:} The regular expression \\\verb|(\d{1,3}\.\d{1,3}\.\d{1,3}\.\d{1,3})| is used to identify IPv4 addresses in the log messages.
\item \textbf{Username Extraction:} The regular expression \texttt{(?:user=\textbar for user \textbar by user )(\textbackslash w+)} is used to extract usernames from various log message formats.

\end{itemize} 

\item \textbf{ASP Fact Generation:} Iterates through the processed log entries and generates ASP facts corresponding to the defined anomaly patterns. The generated facts include: 
\begin{itemize} 
    \item \texttt{login\_attempt(IP, Timestamp, Status, User).} Represents a login attempt, including the IP address, timestamp, status (failed or success), and username. 
    \item \texttt{session(User, UID, Type, Timestamp).} Represents a user session, including the username, user ID, session type (opened or closed), and timestamp. 
    \item \texttt{network\_connection(IP, Timestamp).} Represents a network connection from a specific IP address at a given timestamp. 
    \item \texttt{account\_action(Action, User, Timestamp).} Represents an account-related action (e.g., \texttt{create\_admin}, \texttt{reactivate}) performed by a user at a given timestamp. 
    \item \texttt{klogind\_auth\_failure(IP, Timestamp).} Represents a \texttt{klogind} authentication failure from a specific IP address at a given timestamp. 
    \item \texttt{gdm\_auth\_failure(Timestamp).} Represents a GDM authentication failure at a given timestamp. 
    \item \texttt{logrotate\_abnormal\_exit(Timestamp).} Represents an abnormal exit of the \texttt{logrotate} process at a given timestamp. 
    \item \texttt{named\_soa\_error(Timestamp).} Represents a DNS SOA error at a given timestamp. 
    \item \texttt{xinetd\_connection\_reset(Timestamp).} Represents a xinetd connection reset at a given timestamp. 
    \end{itemize} 
    These facts are then written to an output file (\texttt{facts.lp}) for use by the ASP solver. \end{enumerate}

This preprocessing step is crucial for transforming the unstructured log data into a structured format that can be effectively analyzed using ASP. The regular expressions used for information extraction are designed to be robust and adaptable to different log message formats. However, the accuracy of the anomaly detection system depends on the quality and completeness of the extracted facts.

\subsection{Study Limitations}

We acknowledge several limitations in this proof-of-concept study. The dataset contains 2,000 entries over 263 days (~7.6 entries/day), which is significantly lower than production systems that typically generate thousands of daily log entries. Additionally, the preprocessing uses second-level timestamp precision, which may lose crucial event ordering information when multiple events occur within the same second. The ASP rules were specifically tailored to this Linux log dataset's characteristics and event types, requiring adaptation for different log formats or systems. Despite these constraints, this work demonstrates ASP's applicability to structured log analysis and establishes a foundation for rule-based anomaly detection frameworks.

\subsection{ASP Program Execution}
The ASP program, consisting of the refined anomaly pattern definitions (in \texttt{anomaly\_detection.lp}) and the generated system log facts (in \texttt{facts.lp}), was executed using the Clingo ASP solver (version 5.7.1). The command used for execution was:
\begin{verbatim}
clingo 0 anomaly_detection.lp facts.lp
\end{verbatim}
The ``0" argument instructs Clingo to find all answer sets (models). Each answer set represents a consistent set of detected anomalies based on the rules and facts.

\section{Result Analysis}
The execution of the ASP program on the 
dataset yielded one answer set containing several detected anomalies. Due to the volume of individual brute-force and privilege escalation events, a summary of the types of anomalies found is presented below, along with illustrative examples.

The Clingo output included instances of the following anomalies:
\begin{itemize}
    \item \textbf{network\_anomaly(IP)}: Indicates an IP address with a high frequency of network connections.
        Example: \texttt{network\_anomaly("208.62.55.75")},\\ \texttt{network\_anomaly("207.30.238.8")}. These IPs were observed making numerous connections, potentially indicative of scanning or automated activity.
    \item \textbf{potential\_brute\_force(IP, Timestamp)}: Signifies multiple failed login attempts from a specific IP address around a given timestamp.
        Example: \texttt{potential\_brute\_force("150.183.249.110", 1752163307)}. Many such instances were detected for various IPs, including ``unknown" where the IP was not parsable from the log line.
    \item \textbf{potential\_privilege\_escalation(User, Timestamp)}: Flags instances where a non-root user (e.g., ``cyrus", ``news") opened a session with root privileges (UID 0).
        Example: \texttt{potential\_privilege\_escalation("cyrus", 1749960378)}. This was a common occurrence for system users like ``cyrus" and ``news", likely corresponding to scheduled tasks or administrative scripts run with elevated privileges via ``su".
    \item \textbf{system\_issue(Type, Timestamp)}: Captures various system-level operational issues.
    \begin{itemize}
        \item \texttt{system\_issue(logrotate\_abnormal\_exit, Timestamp)}: \\
        Example: \texttt{system\_issue(logrotate\_abnormal\_exit, 1749960380)}. These indicate that the log rotation process did not complete successfully.
        \item \texttt{system\_issue(named\_soa\_error, Timestamp)}: \\
        Example: \texttt{system\_issue(named\_soa\_error, 1753445346)}. These point to errors related to DNS Start of Authority records.
        \item \texttt{system\_issue(xinetd\_connection\_reset, Timestamp)}: \\
        Example: \texttt{system\_issue(xinetd\_connection\_reset, 1753485793)}. These suggest network connections managed by \texttt{xinetd} were unexpectedly reset.
    \end{itemize}
\end{itemize}

Anomalies related to \texttt{account\_anomaly}, \texttt{potential\_klogind\_brute\_force}, and \texttt{potential\_gdm\_brute\_force} were not present in the output. This suggests that either the specific conditions for these rules (e.g., creation of admin accounts, sufficient \texttt{klogind/GDM} failures to meet thresholds) were not met in the dataset, or the necessary base facts (like \texttt{dormant\_account}) were not available.

The results demonstrate the ASP system's capability to identify a diverse set of predefined anomalous patterns from a real-world, structured log dataset. The explainability of ASP allows for a clear understanding of why each event was flagged, based on the logical rules.

\section{Discussion}

The application of Answer Set Programming (ASP) to detect anomalies in system logs represents an innovative approach in the field of computer forensics and cyber security. The results demonstrate ASP's capability to identify diverse predefined anomalous patterns from real-world, structured log data. The explainability of ASP allows for clear understanding of why each event was flagged, based on the logical rules defined in Section 2.3.

Key advantages observed in this implementation include the declarative rule specification that enabled security experts to easily encode complex detection patterns, the system's ability to handle incomplete log information through default reasoning, and the transparent logical reasoning chains that facilitate security analyst investigations.

\section{Limitations and Future Directions}

While the application of ASP to the 
dataset demonstrates its potential for analyzing real-world system logs, this study also highlights several limitations and areas for future research.

\subsection{Limitations}

\textbf{Dataset Specificity:} The refined anomaly patterns and ASP rules were tailored to the specific characteristics of the 
Linux logs
dataset\cite{zhu2023loghub}. The limited scope of the dataset (e.g., lack of detailed file operation types, process paths, or network traffic volume/protocol details for all connections) restricted the types of anomalies that could be effectively detected. Applying the same rules to different datasets with different log formats or event types may require significant modifications.

\textbf{Contextual Information:} The current approach relies primarily on information extracted directly from the log entries. Incorporating more contextual information, such as network topology, user roles, and system configuration, could significantly improve the accuracy and reduce the number of false positives. For example, frequent privilege escalations for certain service accounts might be normal behavior and should not be flagged as anomalies.

\textbf{Threshold Selection:} The performance of the anomaly detection system is highly sensitive to the choice of thresholds (e.g., \texttt{login\_time\_window}, \texttt{failed\_lo-\\gin\_threshold}, \texttt{connection\_threshold}). Manually tuning these thresholds can be time-consuming and may not generalize well to different datasets or system configurations.

\textbf{Lack of Ground Truth:} The Linux logs
dataset\cite{zhu2023loghub} does not provide ground truth labels indicating which log entries correspond to actual security incidents. This makes it difficult to quantitatively evaluate the performance of the anomaly detection system and compare it with other approaches.

\textbf{Limited Anomaly Types:} The current implementation focuses on a limited set of predefined anomaly patterns. It may not be able to detect novel or unknown types of attacks that do not match these patterns.

\subsection{Future Directions}

\textbf{Machine Learning Integration:} Combining ASP with machine learning techniques could lead to more adaptive and robust anomaly detection systems. Machine learning could be used to:
\begin{itemize}
    \item \textbf{Dynamically Adjust Thresholds:} Train machine learning models to predict optimal threshold values based on the characteristics of the log data and the system environment.
    \item \textbf{Suggest New Rules:} Use machine learning to identify new patterns of anomalous behavior and automatically generate corresponding ASP rules.
    \item \textbf{Improve Fact Extraction:} Train machine learning models to more accurately extract information from unstructured log messages, particularly for fields that are difficult to parse using regular expressions.
\end{itemize}

\textbf{Real-time Processing:} Developing techniques for incremental ASP solving could enable real-time processing of log streams, allowing for immediate detection and response to anomalies. This would require efficient methods for updating the ASP program with new facts and retracting old facts as new log entries arrive.

\textbf{Probabilistic Extensions:} Incorporating probabilistic reasoning into ASP could help in dealing with uncertainties in log data and provide more nuanced risk assessments. This could involve assigning probabilities to different facts and rules based on their reliability and confidence.

\textbf{Automated Rule Generation:} Developing methods to automatically generate or suggest ASP rules from threat intelligence feeds could significantly reduce the manual effort required to keep the system up-to-date. This could involve using natural language processing techniques to extract security rules from threat reports and translate them into ASP predicates.

\textbf{Contextual Enrichment:} Enriching the log data with contextual information from external sources (e.g., network topology, user roles, system configuration) could significantly improve the accuracy and reduce the number of false positives. This could involve integrating the ASP-based system with other security tools and data sources.

\textbf{Evaluation Metrics and Benchmarking:} Developing standardized evaluation metrics and benchmark datasets for ASP-based anomaly detection systems would facilitate the comparison of different approaches and promote the development of more effective techniques.

\textbf{Hybrid Approaches:} 
Exploring hybrid approaches that combine the strengths of logical reasoning with other computational techniques, such as statistical methods and machine learning, could lead to more powerful and versatile anomaly detection systems.

\section{Conclusion}

The application of ASP to the Linux system log analysis demonstrates its potential for cyber intelligence and computer forensics. The framework successfully identified various predefined anomalies, showcasing ASP's strengths in logical reasoning and declarative rule specification. However, challenges in rule refinement, context incorporation (e.g., distinguishing malicious from benign privilege escalations), and handling dataset-specific limitations (e.g., missing fields for some desired anomaly patterns) highlight areas for further work.

As the field of cybersecurity continues to evolve, ASP-based approaches, particularly when combined with data preprocessing and potentially machine learning for dynamic rule adjustment, have the potential to play a significant role in next-generation intrusion detection and log analysis systems. By addressing current limitations and exploring integration with other technologies, ASP can become an even more valuable tool in the cybersecurity arsenal.

\bibliographystyle{splncs04}
\bibliography{references}

\end{document}